\documentclass[a4paper,12pt]{article}
\usepackage{tabularx}
\usepackage{amsmath}
\usepackage{amssymb}
\usepackage{graphicx}
\usepackage[margin=0.7in,a4paper]{geometry}
\usepackage[section]{placeins}
\usepackage{mathcomp}
\usepackage{authblk}
\usepackage{url}

\usepackage{gensymb}
\usepackage{float}

\title{\vspace{-1cm} Tuning electronic and optical properties of 2D polymeric C$_{60}$ by stacking two layers}

\author[$\dagger$,1]{Dylan Shearsby}
\affil[$\dagger$]{These authors contributed equally to this work.}
\affil[1]{Clare College, University of Cambridge, Trinity Lane, Cambridge CB2 1TL, United Kingdom}
\author[$\dagger$,2]{Jiaqi Wu}
\affil[2]{Peterhouse, University of Cambridge, Trumpington Street, Cambridge CB2 1RD, United Kingdom}
\author[1]{Dekun Yang}
\author[3,*]{Bo Peng}
\affil[3]{Theory of Condensed Matter Group, Cavendish Laboratory, University of Cambridge, J.\,J.\,Thomson Avenue, Cambridge CB3 0HE, United Kingdom}
\affil[*]{bp432@cam.ac.uk}
\date{\vspace{-5ex}}

\begin{document}

\maketitle

\begin{abstract}
Benefiting from improved stability due to stronger interlayer van der Waals interactions, few-layer fullerene networks are experimentally more accessible compared to monolayer polymeric C$_{60}$. However, there is a lack of systematic theoretical studies on the material properties of few-layer C$_{60}$ networks. Here, we compare the structural, electronic and optical properties of bilayer and monolayer fullerene networks. The band gap and band-edge positions remain mostly unchanged after stacking two layers into a bilayer, enabling the bilayer to be almost as efficient a photocatalyst as the monolayer. The effective mass ratio along different directions is varied for conduction band states due to interlayer interactions,
leading to enhanced anisotropy in carrier transport. Additionally, stronger exciton absorption is found in the bilayer than that in the monolayer over the entire visible light range, rendering the bilayer a more promising candidate for photovoltaics. Moreoever, the polarisation dependence of optical absorption in the bilayer is increased in the red-yellow light range, offering unique opportunities in photonics and display technologies with tailored optical properties over specific directions. Our study provides strategies to tune electronic and optical properties of 2D polymeric C$_{60}$ via the introduction of stacking degrees of freedom.
\end{abstract}

\section{Introduction}

Recently, monolayer fullerene networks have been synthesised through organic cation slicing exfoliation\,\cite{Hou2022}. This combines the characteristics of 0D fullerene molecules and 2D graphene monolayers such as large surface areas and high flexibility, making it promising for applications such as photocatalytic water splitting\,\cite{Peng2022c,Jones2023,Wu2024a,Tromer2022}. However, achieving stable monolayer fullerenes remains challenging due to their inherent conflicting dynamic, thermodynamic and mechanical stability\,\cite{Peng2023,Ribeiro2022}. This results in C$_{60}$ monolayers being prone to degradation into 1D or 0D at high temperatures. As a result, it has been found that mechanical exfoliation of van der Waals polymeric C$_{60}$ can only lead to few-layer covalent network of fullerenes instead of the single-layer form\,\cite{Meirzadeh2023}. The obtained few-layer fullerene networks, consisting of two or more stacked layers, have been more commonly produced and studied experimentally\,\cite{Meirzadeh2023,Wang2023}. Few-layer C$_{60}$ networks benefit from enhanced structural stability due to stronger interlayer van der Waals interactions. This provides more stable configurations in multiple structural phases\,\cite{Hou2022,Meirzadeh2023} while still retaining many of the desirable chemical properties predicted in monolayers such as photocatalysis\,\cite{Wang2023}.

\begin{figure}[ht]
    \centering
    \includegraphics[width=0.5\linewidth]{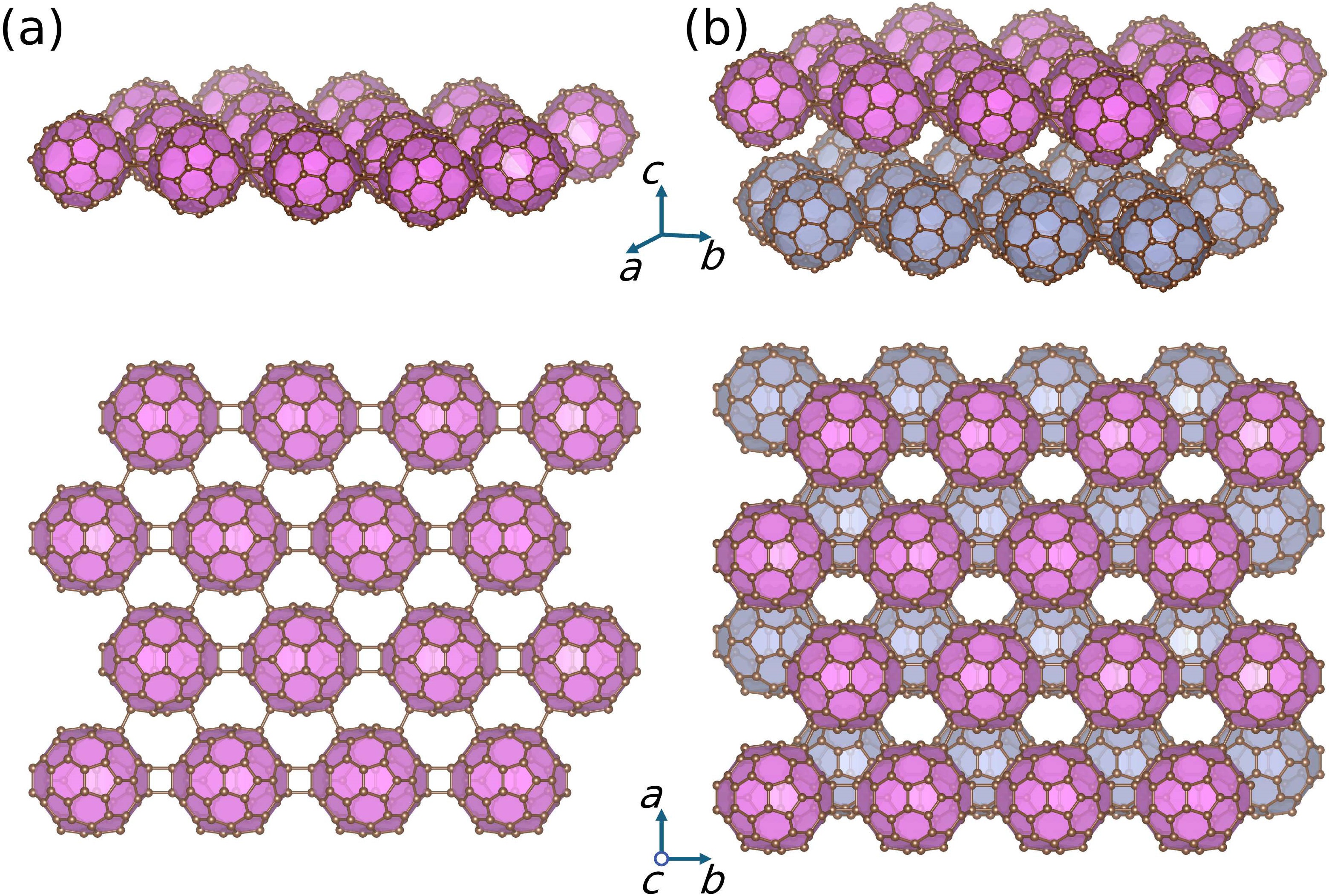}
    \caption{3D and top views of the crystal structures of (a) monolayer and (b) bilayer fullerene networks visualised by {\sc vesta}\,\cite{vesta}. 
    }
    \label{fig:crystal}
\end{figure} 

Bilayer polymeric C$_{60}$ represents the simplest form of few-layer fullerene networks, with the two layers stacked together as shown in Fig.\,\ref{fig:crystal}. The bilayer structure introduces additional degrees of freedom for tailored function by varying their stacking orders such as orientation, sliding, and twisting angle. Such structural tunability allows control over electronic and optical properties such as effective masses for carrier transport and anisotropic optical absorption for polarised light detectors\,\cite{Peng2016d}, making these materials ideal candidates for use in flexible electronics, photodetectors, and optoelectronics. Additionally, the capability to tune interlayer sliding or twisting angles can lead to emergent phenomena, such as sliding ferroelectricity\,\cite{Wang2024a} or moir{\'e} patterns\,\cite{Meirzadeh2023}. These unique properties suggest bilayer fullerene networks can also be applied in sensors, capacitors, and memory storage devices. However, there is a lack of theoretical investigations on the structural, electronic and optical properties of bilayer fullerene networks for a range of applications from flexible electronics to photovoltaics.

In this paper, we delve into the electronic structures and excitonic effects of bilayer C$_{60}$ networks whilst comparing to the monolayer C$_{60}$ network, using state of-the-art first principles calculations. By using the experimentally reported AB stacked bilayer structure, we assess the electronic properties including band gap, band-edge positions and carrier effective mass based on unscreened hybrid functional calculations. The optical properties are then investigated for the potential applications in photonics and optoelectronics.

\section{Methods}\label{Methods}

Density functional theory (DFT) calculations were performed using the Vienna \textit{ab initio} Simulation Package ({\sc VASP})\,\cite{Kresse1996,Kresse1996a}. The projector-augmented wave basis set\,\cite{Bloechl1994,Kresse1999} was employed with the Perdew-Burke-Ernzerhof functional revised for solids (PBEsol)\,\cite{Perdew2008} within the generalised gradient approximation (GGA). A plane-wave cutoff energy of 800\,eV was used with the Brillouin zone sampled using a $\Gamma$-centred \textbf{k}-point grid of $2\times3$. The zero damping DFT-D3 method of Grimme was included to described the van der Waals interactions\,\cite{Grimme2010}. A vacuum spacing larger than 24\,\AA\ was applied with the dipole corrections along $z$\,\cite{Makov1995}. The crystal structures were fully relaxed using the energy and force convergence criteria of $10^{-6}$\,eV and $10^{-2}$\,eV/\AA\ respectively. For electronic structures, hybrid functional calculations was used by mixing 75\% of the PBEsol exchange functional with 25\% unscreened exact Hartree-Fock exchange\,\cite{Adamo1999}, denoted as PBEsol0. The effective masses at the band edges were calculated by sampling the Brillouin Zone near $\Gamma$ in intervals of 0.015\,\AA$^{-1}$ with 5 points each along $\Gamma - \mathrm{X}$ and $\Gamma - \mathrm{Y}$, and then fitting the energy dispersion curves with a cubic polynomial. Excitonic calculations were performed using the time-dependent Hartree-Fock (TDHF) method with the Casida equation\,\cite{Sander2017} from a basis of the 16 highest valance bands and the 16 lowest conduction bands with a well converged $\mathbf{k}$-mesh of $4\times6$ for the monolayer and $2\times3$ for the bilayer. The TDHF approach is similar to the Bethe-Salpeter equation (BSE) but crucially uses the exchange-correlation kernel for the screening of the Coulomb potential instead of the screened exchange $W$, yielding quantitatively agreeable excitonic properties\,\cite{Peng2022c,Jones2023} with the computationally heavy $GW$\,+\,BSE results\,\cite{Champagne2024}.

\section{Results and Discussion}\label{Results}

\subsection{Crystal structures}

Figure\,\ref{fig:crystal} shows the crystal structures of monolayer and bilayer C$_{60}$ networks. For the monolayer, each carbon cage is connected by the C$-$C single bonds along the [110] direction and the in-plane [$2+2$] cycloaddition bonds along the [010] direction, leading to a quasi-hexagonal lattice with space group $Pmna$ (No.\,53). Bilayer C$_{60}$ networks have a space group $P2/c$ (No.\,13) and can be viewed as an AB stacking of two layers held together by the van der Waals interactions. The AB stacking pattern leads to a closely-packed lattice as expected for nearly-spherical C$_{60}$ molecules to be arranged in a space-efficient manner, as reported experimentally\,\cite{Meirzadeh2023}. The lattice constants of the bilayer $a = 15.804$\,\AA\ and $b = 9.115$\,\AA\ are slightly larger than those of the monolayer $a = 15.798$\,\AA\ and $b = 9.110$\,\AA, agreeing well with the synthesised structures\,\cite{Hou2022,Meirzadeh2023}.

\begin{table}[ht]
    \caption{
    Lattice constants of bilayer fullerene networks from different functionals, as well as the layer distance $\Delta h$ (defined as the height difference between the lowest atom in top layer and the highest atom in bottom layer) and the actual distance $\Delta d$ between the corresponding atoms.}
    \centering
    \begin{tabular}{ccccc}
    \hline
     & $a$ (\AA) & $b$ (\AA) & $\Delta h$ (\AA) & $\Delta d$ (\AA) \\ \hline
    PBE & 15.931 & 9.171 & 2.251 & 5.962 \\
    PBE\,+\,D3 & 15.889 & 9.157 & 1.662 & 5.763 \\
    PBEsol & 15.840 & 9.128 & 1.912 & 5.885 \\
    PBEsol\,+\,D3 & 15.804 & 9.115 & 1.669 & 5.840 \\ \hline
    \end{tabular}  
    \label{tbl:abc} 
\end{table}

\subsection{Band structures}

\begin{figure}[ht]
    \centering
    \includegraphics[width=\linewidth]{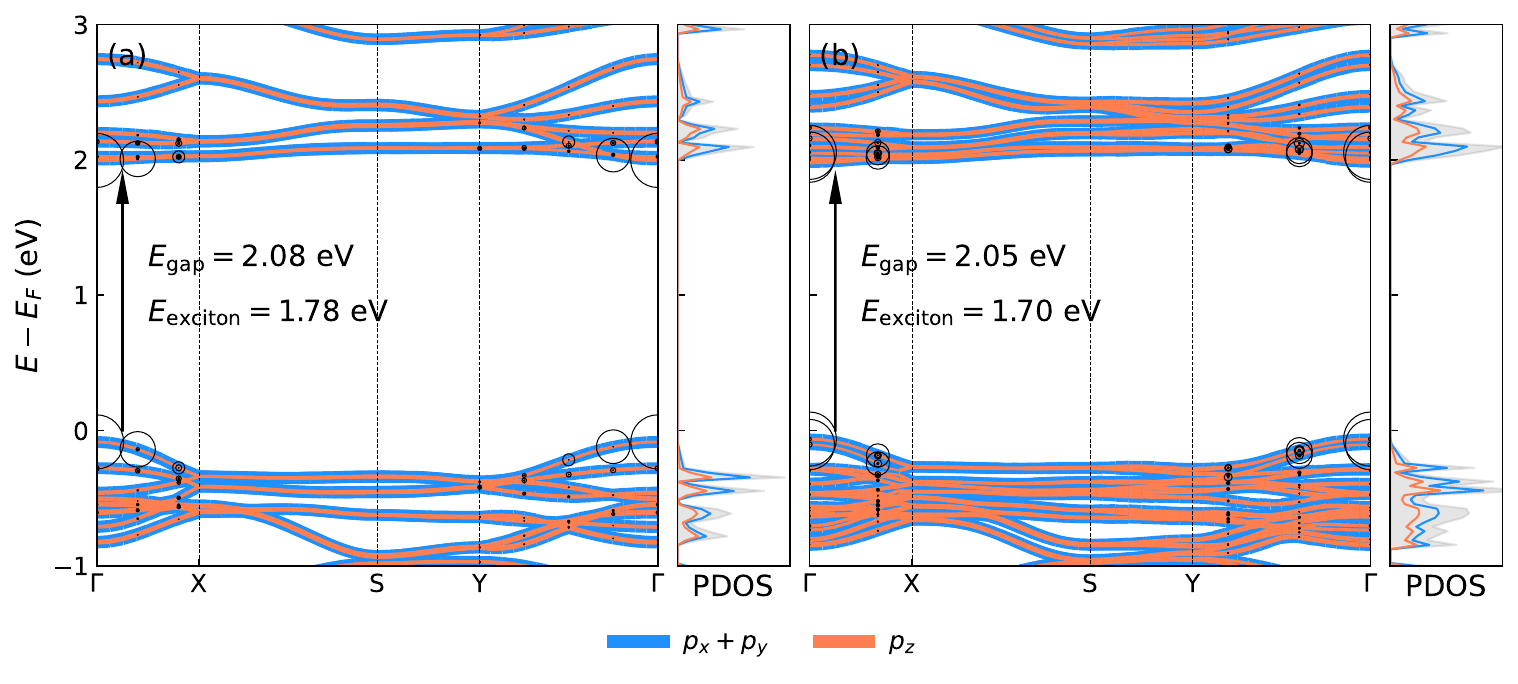}
    \caption{Orbital-projected band structures of (a) monolayer and (b) bilayer fullerene networks at PBEsol0 level. The circles show the contributions of each electron-hole pair to the lowest energy bright excitons.}
    \label{fig:excitons}
\end{figure}

The PBEsol0 band structures of monolayer and bilayer fullerene networks are shown in Fig.\,\ref{fig:excitons}. Both structures show a direct band gap at $\Gamma$ with the bilayer gap of 2.05\,eV slightly lower than the monolayer gap of 2.08\,eV, which is due to the weak interlayer interactions in the bilayer. As shown in Table\,\ref{tbl:abc}, despite that the layer distance $\Delta h$ between the lowest atom in top layer and the highest atom in bottom layer is quite small ($1.662-2.251$\,\AA\ for different functionals), the actual distance between the corresponding atoms $\Delta d$ is much larger (> 5.840\,\AA) because of the AB stacking, leading to the weak interlayer interactions.

\begin{figure*}[ht]
    \centering
    \includegraphics[width=0.85\linewidth]{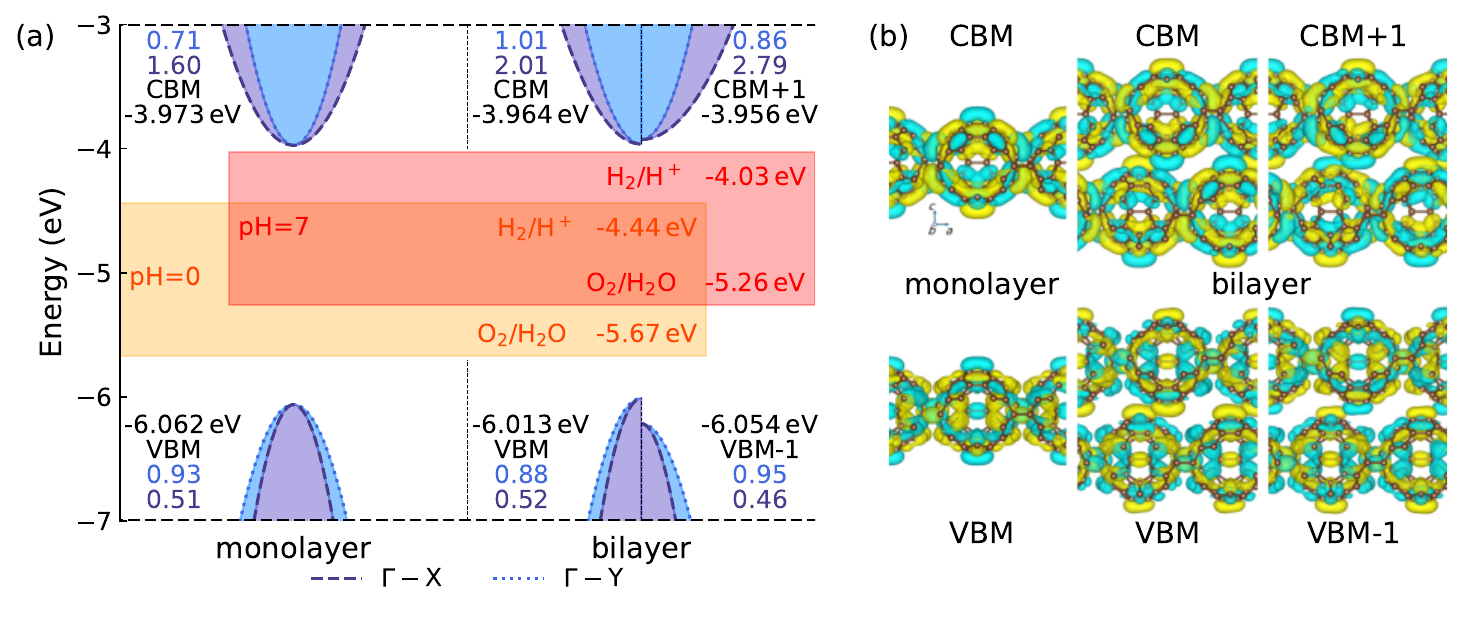}
    \caption{(a) Band alignment of monolayer and bilayer fullerene networks with the effective masses $m_{\mathrm{e}}$ labelled for the band edges. The water redox potentials are also marked. The eigenenergy difference between bilayer CBM and CBM+1, as well as between VBM and VBM-1, are zoomed in $\times5$. (b) Side view of the band-edge states for monolayer and bilayer fullerene networks. 
    }
    \label{fig:photocatalysis}
\end{figure*}

The computed band gap for the bilayer agrees well with the measured gap for few-layer polymeric C$_{60}$ of 2.02\,\cite{Meirzadeh2023} and 
2.05\,eV\,\cite{Wang2023}, while the monolayer gap from unscreened hybrid functional PBEsol0 is comparable with the gap from the many-body perturbation theory\,\cite{Champagne2024}. Because the band gap of bilayer polymeric C$_{60}$ is similar to that of the monolayer, the energy levels of their conduction band minimum (CBM) and valence band maximum (VBM) straddle the redox potentials of water in Fig.\,\ref{fig:photocatalysis}(a), similar to monolayer fullerene networks\,\cite{Peng2022c}. As a result, the band edges of bilayer C$_{60}$ networks provide sufficient driving force for overall water splitting, as observed experimentally\,\cite{Wang2023}.
In bilayer fullerene networks, there are two nearly degenerate VBM states and two nearly degenerate CBM states, denoted as VBM-1/VBM and CBM/CBM+1 respectively (labelling from low to high eigenenergies). The nearly degenerate VBM-1 and VBM (CBM and CBM+1) states can be understood as the in-phase and out-of-phase combinations of the VBM (CBM) states of the top and bottom layers that make up the bilayer, as demonstrated in Fig.\,\ref{fig:photocatalysis}(b). This suggests similar orbital characters on the top and bottom layers, preserving the band-edge positions and hence the catalytic reactivity of the monolayers. The VBM-1 and VBM (or CBM and CBM+1) states have slightly different eigenenergies because of the interlayer interactions\,\cite{bian_strong_2022}.

We then compare the effective masses of the two systems in Fig.\,\ref{fig:photocatalysis}(a). For monolayer fullerene, the effective masses of holes are smaller than those of electrons. The holes along $\Gamma - \mathrm{X}$ have the smallest effective mass of 0.51\,$m_\mathrm{e}$, while the electrons along $\Gamma - \mathrm{X}$ possess the largest effective mass of 1.60\,$m_\mathrm{e}$. This is in good agreement with the transport calculations showing that the holes along $a$ have the largest mobility $\ge$\,70\,cm$^2$/Vs and the electron mobility along $a$ is below 5\,cm$^2$/Vs.

For the bilayer, the hole effective masses of the VBM-1 (VBM) states are 0.52\,$m_\mathrm{e}$ (0.46\,$m_\mathrm{e}$) along $\Gamma - \mathrm{X}$ and 0.88\,$m_\mathrm{e}$ (0.95\,$m_\mathrm{e}$) along $\Gamma - \mathrm{Y}$ respectively, with both directions having comparable results to the monolayer, i.e., 0.51\,$m_\mathrm{e}$ and 0.93\,$m_\mathrm{e}$ along $\Gamma - \mathrm{X}$ and $\Gamma - \mathrm{Y}$ respectively. On the other hand, the conduction bands of the bilayer have larger effective mass than the monolayer, with the largest effective mass of 2.79\,$m_\mathrm{e}$ in CBM+1 along $\Gamma - \mathrm{X}$. Moreover, the ratios between electron effective masses along $\Gamma - \mathrm{X}$ and $\Gamma - \mathrm{Y}$ are 2.25, 1.99 and 3.24 for monolayer, CBM and CBM+1 respectively, indicating a change in transport anisotropy.  
The directional dependence of electron mobility provides unique advantages in designing high-performance organic transistors and flexible electronics with tailored electron conductivity.

\subsection{Optical properties}

\begin{figure}[ht]
    \centering
    \includegraphics[width=0.85\linewidth]{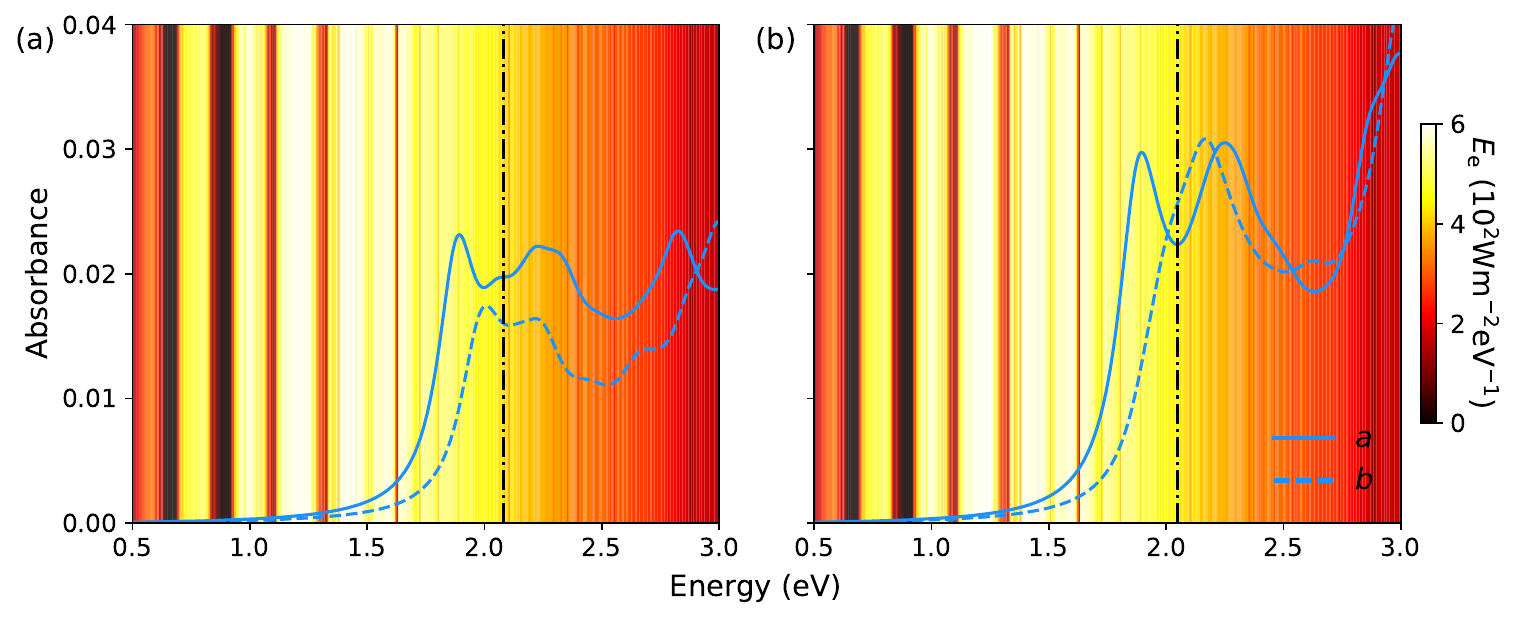}
    \caption{Optical absorbance of (a) monolayer and (b) bilayer fullerene networks for both $a$ and $b$ polarisations. The black vertical dash-dot lines represent the direct band gaps from independent-particle approximations in the absence of excitons. The background shows the global total spectral irradiance $E_{\rm e}$ from the Sun.}
    \label{fig:optical}
\end{figure}

For the monolayer, the first bright exciton mainly consists of the electron-hole pair around the CBM and VBM at $\Gamma$ with an exciton eigenenergy of 1.78\,eV, leading to a high binding energy of 0.30\,eV. For the bilayer, the first bright exciton is mostly contributed by holes from both VBM-1 and VBM and electrons from both CBM and CBM+1, leading to a lower exciton energy energy of 1.70\,eV. Consequently, the exciton binding energy of the bilayer becomes slightly larger (0.35\,eV).

Finally, we compare the optical absorbance of the two systems as seen in Fig.\,\ref{fig:optical}. Both the monolayer and bilayer systems exhibit strongly bound excitons that dominate the optical absorption in the red-yellow light range below the band gap. For the monolayer, the absorbance for polarisation along $a$ and $b$ is slightly anisotropic. Strong exciton absorption peaks can be found along both polarisation directions below the band gap. Stacking two monolayers together results in enhanced absorbance as expected. Moreover, the anisotropy in absorbance along $a$ and $b$ polarisation directions also becomes stronger: a strong exciton absorption peak appears below the band gap for polarisation along $a$, while the absorption peak along $b$ occurs above the band gap. 

The optical absorbance for both systems exhibit strong absorption peaks over the entire visible light range, making them promising for photovoltaics. Additionally, the strong anisotropic absorbance provides opportunities for advanced photonic devices such as polarised light detectors. The optical properties can be further tuned by manipulating layer orientations, offering distinct advantages in display technologies and flexible electronics. In display technologies, anisotropic optical absorption allows colour tuning over specific polarisation directions with reduced power consumption. Additionally, strong absorption in specific directions enhances contrast and clarity to improve overall display quality. For flexible electronics, the tunable optical properties of bilayer C$_{60}$ networks align with the demand for flexible materials that can efficiently absorb and manipulate light.

\section{Conclusions}
To conclude, we investigate the structural, electronic and optical properties of bilayer fullerene networks in comparison with their monolayer counterpart. After stacking two layers into a bilayer, the positions of band edges only slightly change and remain suitable for photocatalytic water splitting. The electron mobility of bilayer polymeric C$_{60}$ shows varied anisotropy due to the interlayer interactions of electron wavefunctions. 
The optical absorbance of the bilayer is enhanced over the entire visible light range compared to the monolayer, offering the opportunity for organic solar cells with higher efficiency. Additionally, the polarisation dependence of optical absorbance in the bilayer is stronger than that of the monolayer, leading to unique advantages in photonic devices. The combination of promising optical properties and flexibility 
makes bilayer C$_{60}$ networks specifically useful for next-generation displays and wearable electronic devices. Our work demonstrates the potential to manipulate materials properties of 2D fullerene networks by stacking individual layers into bilayers, opening a promising route to further exploration of stacking degrees of freedom such as orientations, sliding, and twisting angles.

\section*{Acknowledgements}
J.W. acknowledges support from the Cambridge Undergraduate Research Opportunities Programme and from Peterhouse for a Bruckmann Fund grant and the James Porter Scholarship. D.Y. acknowledges support from Clare College for the Undergraduate Travel \& Research Grant. B.P. acknowledges support from Magdalene College Cambridge for a Nevile Research Fellowship. The calculations were performed using resources provided by the Cambridge Service for Data Driven Discovery (CSD3) operated by the University of Cambridge Research Computing Service (\url{www.csd3.cam.ac.uk}), provided by Dell EMC and Intel using Tier-2 funding from the Engineering and Physical Sciences Research Council (capital grant EP/T022159/1), and DiRAC funding from the Science and Technology Facilities Council (\url{http://www.dirac.ac.uk} ), as well as with computational support from the UK Materials and Molecular Modelling Hub, which is partially funded by EPSRC (EP/T022213/1, EP/W032260/1 and EP/P020194/1), for which access was obtained via the UKCP consortium and funded by EPSRC grant ref EP/P022561/1.

\end{document}